# Experimental Determination of the Lorenz Number in $Cu_{0.01}Bi_2Te_{2.7}Se_{0.3}$ and $Bi_{0.88}Sb_{0.12}$

K. C. Lukas,[1] W. S. Liu,[1] G. Joshi,[1] M. Zebarjadi,[3] M. S. Dresselhaus,[2] Z. F. Ren,[1] G. Chen,[3] C. P. Opeil[1]

[1]Department of Physics, Boston College, Chestnut Hill, MA 02467, USA

[2]Department of Physics and Department of Electrical Engineering and Computer Science, Massachusetts Institute of Technology, Cambridge, MA 02139, USA

[3]Department of Mechanical Engineering and Computer Science, Massachusetts Institute of Technology, Cambridge, MA 02139, USA

ABSTRACT

Nanostructuring has been shown to be an effective approach to reduce the lattice thermal conductivity and improve the thermoelectric figure of merit. Because the experimentally measured thermal conductivity includes contributions from both carriers and phonons, separating out the phonon contribution has been difficult and is mostly based on estimating the electronic contributions using the Wiedemann-Franz law. In this paper, an experimental method to directly measure electronic contributions to the thermal conductivity is presented and applied to $Cu_{0.01}Bi_2Te_{2.7}Se_{0.3}$, $[Cu_{0.01}Bi_2Te_{2.7}Se_{0.3}]_{0.98}Ni_{0.02}$, and $Bi_{0.88}Sb_{0.12}$. By measuring the thermal conductivity under magnetic field, electronic contributions to thermal conductivity can be extracted, leading to knowledge of the Lorenz number in thermoelectric materials.

I. INTRODUCTION

The determination of the Lorenz number is an important aspect in thermoelectric research due to the fact that ZT enhancement is being realized through the reduction of thermal conductivity, specifically focusing on reducing the lattice portion of the thermal conductivity. The total thermal conductivity is given by

$$\kappa_{total} = \kappa_{carrier} + \kappa_{lattice} \qquad (1)$$

where $\kappa_{carrier}$ and $\kappa_{lattice}$ are the contributions to the thermal conductivity from the carriers and the lattice, respectively. Since only the total thermal conductivity can be measured, the contributions must be separated in some way. This is done using the Wiedemann-Franz Law and by defining a Lorenz number (L), which is the given by

$$L = \frac{\kappa_{carrier}}{\sigma T} \qquad (2)$$

where $\sigma$ is the electrical conductivity and T is the absolute temperature. In metals the Lorenz number can be determined by measuring the electrical conductivity and total thermal conductivity at a given temperature, from which the Lorenz number is calculated using equation 2. This method is only useful in metals where the total thermal conductivity is approximately equal to $\kappa_{carrier}$. For the classical free electron model the Lorenz number is given as $2.44\times10^{-8}$ $V^2K^{-2}$.[1] It is important to note that the Lorenz number, as described by the free electron model, is not an accurate value for most materials and in a given material depends on the detailed band structure, position of the Fermi level, and the temperature; for semiconductors this relates to the carrier concentration. Therefore, when $\kappa_{lattice}$ and $\kappa_{carrier}$ become comparable to each other, there must be a method for differentiating between the two components of $\kappa_{total}$. To date the separation of the two components has been accomplished through calculation by approximating the Lorenz number, and hence the carrier contribution, through various different formalisms.[1,3-4] Determinations of the Lorenz number have also been made experimentally,[1] however there are few.

In order to separate $\kappa_{lattice}$ and $\kappa_{carrier}$ experimentally, two approaches have been used to determine the Lorenz number. Both methods utilize a transverse magnetic field in order to suppress the electronic component of the thermal conductivity. One approach uses a classically large magnetic field while the other is performed in intermediate fields. A classically large magnetic field is described as $\mu B >> 1$ where $\mu$ is the carrier mobility and B is the magnetic field.[1] When this limit is reached, the electronic component of $\kappa$ is completely suppressed so that the measurement yields only the lattice portion of the thermal conductivity, from which $\kappa_{carrier}$ and hence the Lorenz number can be calculated using equations 1 and 2.

Very often it is difficult to reach a classically large field, making this type of measurement not possible and therefore other methods have been developed for determining L. For example, Goldsmid et al. developed a Magneto-Thermal Resistance (MTR) method for extracting the Lorenz number at lower magnetic fields, specifically in the region where $\mu B \approx 1$.[5-8] In the MTR method the sample is kept at a constant temperature while the field is varied. In this case both the electrical conductivity as well as the total thermal conductivity will change with the field due to the Lorentz force acting on the carriers, which is induced by the transverse magnetic field. Equation 1 can be rewritten in the form

$$\kappa(B)_{total} = LT\sigma(B) + \kappa_{lattice} \qquad (3)$$

where now both $\kappa$ and $\sigma$ are dependent on magnetic field. It is noted that $\kappa$, $\sigma$ and L are all tensors, whose off diagonal components can have a non-negligible contributions in magnetic field.[5,9] Both $\kappa$(B) and $\sigma$(B) are measured along the same direction, which we define as $\kappa_{xx}(B)$ and $\sigma_{xx}(B)$. For an anisotropic sample, even to the first order, the magnetic field affects the diagonal terms of the tensors as well as the non-diagonal terms. We show that by measuring only the diagonal terms we are able to extract the Lorentz number $L_{xx}$, which relates the $\kappa_{xx}$ to $\sigma_{xx}$. The reason behind the validity of this method is that both $\kappa(B)_{xx}$ and $\sigma(B)_{xx}$ have a similar magnetic field dependence and their ratio has only a weak dependence on the off-diagonal terms. Since the samples are isotropic[10] and extrinsic it is assumed that off diagonal terms can be neglected because thermogalvanomagnetic effects are only dominant in intrinsic materials with a proportional number of positive and negative charge carriers.[4,11] As long as both have the same functional form with respect to the magnetic field, then $\kappa$(B) vs. $\sigma$(B) will have a linear relationship and the Lorenz number $L_{xx}$ at a given temperature can be directly taken from the slope as given in equation 3. It is important to note that the analysis throughout this paper is based on the assumption that the Lorenz number is independent of magnetic field, which is true for some materials but in general is not a valid assumption.[12-14] Analogous approximations have been used to study similar compounds in the past.[6,12]

Neither method has been extensively used due to the fact that there are restrictions on the materials that can be measured because there must be a significant carrier contribution to the total thermal conductivity; also the experimental setup is rather difficult to realize.[1,5-9] The advent of the Physical Properties Measurement System (PPMS) from Quantum Design makes the experimental setup and measurement readily possible for either method. The purpose of this paper is to present experimental techniques for the determination of the Lorenz number from which both the electronic and lattice contributions to the thermal conductivity can be directly extracted. Measurements are compared to literature values as well as simple model calculations. There are several different ways to analyze the raw experimental data; two different models will be used here and shown to yield similar results. The measurements are performed below 150 K so that bipolar terms will be negligible and therefore equations 1 and 3 accurately describe the contributions to the total thermal conductivity. While this technique has been used before, to the best of our knowledge this experimental method has not previously been demonstrated on nanostructured thermoelectrics.

## II. EXPERIMENTAL

Samples were prepared by combining the proper stoichiometric ratios of Cu (99.999%, Alfa Aesar), Bi (99.999%, Alfa Aesar), Te (99.999%, Alfa Aesar), and Se (99.999%, Alfa Aesar) for $Cu_{0.01}Bi_2Te_{2.7}Se_{0.3}$, while $Bi_{0.88}Sb_{0.12}$ was prepared with Bi (99.999%, Alfa Aesar) and Sb (99.999%, Alfa Aesar). Samples were then ball milled and pressed using dc hot-pressing techniques.[10] Metallic contacts were sputtered onto the surfaces so that electrical contacts could be soldered to the sample.

MTR measurements were performed using the Thermal Transport Option (TTO) of the PPMS in which the sample was placed in an orientation where the magnetic field was perpendicular to the heat flow. A standard two point method was used for thermal conductivity and Seebeck coefficient (S) measurements with typical sample dimensions of 2x2x3 $mm^3$. In this case the temperature was held constant at 100 K and measurements were made while the field was swept over a range of 0.1 – 5 T. Since resistivity ($\rho$) values in a magnetic field are required, a four point technique must be used which was accomplished with the AC Transport option on a different sample of dimensions 1x2x12 $mm^3$ for the same temperature and field range. Since a four point technique is used, there is no concern of electrical contact resistance. For thermal contact resistance, our previous measurements show no difference in the thermal conductivity when a two or four point method is used. Even so, any thermal contact resistance is assumed to be negligible in field and since we are looking at the change in thermal conductivity with field, there should be no influence on the slope (L) of the measurement. Geometrical effects on the magnetoresistance are considered to be negligible because the sample used for resistivity measurements in field has the appropriate aspect ratio. The sample dimensions for the thermal magnetoresistance measurements are restricted due to requirements to fit into the PPMS, however it is assumed there is a negligible contribution because there was no evidence previously of geometrical effects on a similar material which had an aspect ratio of 1.[12] Error for the MTR measurements of L and $\kappa_{lattice}$ were calculated from the standard deviation and propagation of error, and determined to be 3% and 7% respectively.

When determining the Lorenz number in a classically large field, the Thermal Transport Option of the PPMS in which the magnetic field was perpendicular to the heat flow was again used. A standard two point method was used for all transport measurements on the same sample. The sample was run in magnetic fields of 0, 6, and 9 T. Only the thermal conductivity measurements in field are used, while electrical resistivity values are taken from the zero field data. Typical sample dimensions were 2x4x2 $mm^3$. Thermal contact resistance is assumed to be negligible for the reasons stated above, and electrical contact resistance is negligible from the comparison of two and four point resistivity measurements. There is no concern of geometrical effects on thermal conductivity measurements because saturation would not be obtained at higher magnetic fields. The measurements were performed over a temperature range of 5-150 K, with error for L and $\kappa_{lattice}$ being 2% and 6% respectively determined from the standard deviation and propagation of error.

## III. RESULTS AND DISCUSSION

The MTR approach can be used only if the thermal and electrical conductivities have the same functional form with respect to the magnetic field. Since the MTR method is used in intermediate fields, or when $\mu B \approx 1$, only values in magnetic fields from 0.8 – 5 T were used, anything below 0.8 T was too low of a field. The top left inset in Figure 1 plots $\kappa$ as a function of field while the lower right inset plots $\sigma$ as a function of field for $Cu_{0.01}Bi_2Te_{2.7}Se_{0.3}$. Both the electrical and thermal conductivity vary with field as $\frac{aB^2}{1+cB^2}$ where a and c are constants, which is valid for strong degeneracy.[2,16-17] The fits are shown in the insets of Figure 1 along with the measured values. It is also noted that the plot of $\kappa$(B) vs. $\sigma$(B) in Figure 1 would not be linear if they did not have the same functional dependence on field. Figure 1 can be fit linearly and taking the slope yields LT in equation 3 from which we get L = 2.16x$10^{-8}$ $V^2K^{-2}$ by dividing by T = 100 K. The lattice portion of the thermal conductivity is given by the y-intercept and gives $\kappa_{lattice}$ = 1.49 W/mK. Care should be taken with the determination of $\kappa_{lattice}$ this way because a larger error is induced when extrapolating over six orders of magnitude to get $\kappa_{lattice}$ when $\sigma$(B) is zero. If $\kappa_{carrier}$ is calculated from the Lorenz number and the electrical conductivity in zero field, $\kappa_{lattice}$ can be calculated from $\kappa_{total} - \kappa_{carrier}$ which gives a value of 1.35 W/mK. For a comparison with the measured values, a simple model for the calculation of the Lorenz number is given by[3]

$$L = \left(\frac{k_B}{e}\right)^2 \left[ \frac{(r+7/2)F_{r+5/2}(\xi)}{(r+3/2)F_{r+1/2}(\xi)} - \left[ \frac{(r+5/2)F_{r+3/2}(\xi)}{(r+3/2)F_{r+1/2}(\xi)} \right]^2 \right] \qquad (4)$$

where r is the scattering parameter, $k_B$ is Boltzmann's constant, e is the electron charge, and $F_n(\xi)$ is the Fermi integral given by

$$F_n(\xi) = \int_0^\infty \frac{\chi^n}{1+e^{\chi-\xi}} d\chi \qquad (5)$$

where $\xi$ is the reduced Fermi energy that can be calculated from the Seebeck coefficient (S) as well as the scattering parameter (r) given by

$$S = \pm \frac{k_B}{e} \frac{(r+5/2)F_{r+3/2}(\xi)}{(r+3/2)F_{r+1/2}(\xi)} - \xi \qquad (6)$$

In this model the Lorenz number can be calculated with knowledge of the Seebeck coefficient and the scattering parameter, both of which were measured at 100 K. The scattering parameter (r) was determined by taking the slope of ln($\mu$) vs. ln(T) around 100 K, using the relationship $\mu \propto T^{r-1}$.[15] The calculated value using equations 4-6 gives L = 2.35x$10^{-8}$ $V^2K^{-2}$ and $\kappa_{lattice}$ = 1.30 W/mK, both of which are close to the experimentally determined values.

The same procedure was followed for $[Cu_{0.01}Bi_2Te_{2.7}Se_{0.3}]_{0.98}Ni_{0.02}$ and Figure 2 shows again that $\kappa$(B) vs. $\sigma$(B) is linear. The measured value for the slope gives L = 2.33x$10^{-8}$ $V^2K^{-2}$ and from the y-intercept $\kappa_{lattice}$ = 1.27 W/mK. The calculated values using equations 4-6 give L = 2.38x$10^{-8}$ $V^2K^{-2}$ and $\kappa_{lattice}$ = 1.13 W/mK, again showing the validity of the measurement. Besides the MTR method the data can also be fit using the following expressions for the electrical and thermal conductivities as a function of field for isotropic samples in the relaxation time approximation.[18]

$$\sigma(B) = \frac{\sigma_0}{1+(\mu_d B)^2} \qquad (7)$$

$$\kappa(B) = \kappa_{lattice} + \frac{\kappa_{carrier}}{1+(\mu_d B)^2} \qquad (8)$$

where $\sigma_0$ is the electrical conductivity in zero field and $\mu_d$ is the drift mobility. The drift mobility determined by equation 7 and shown in Figure 3 is used in equation 8 in order to determine the carrier and lattice contributions to the thermal conductivity as shown in Figure 4. As opposed to the MTR method, the data must be fit using both weak and intermediate magnetic fields and so Figures 3 and 4 show the thermal and electrical conductivity in fields of 0.1 – 5 T. Fitting equation 8 to the thermal conductivity versus magnetic field data in Figure 4 yields $\kappa_{lattice}$ = 1.29 W/mK. It can be seen that using a completely different model presented in equations 7 and 8 produces a nearly identical value of $\kappa_{lattice}$ = 1.27 W/mK as determined by the MTR method.

Unlike $Cu_{0.01}Bi_2Te_{2.7}Se_{0.3}$, it was possible to reach the classical field limit at lower temperatures for Bismuth Antimony compounds. Figure 5 plots the thermal conductivity of $Bi_{0.88}Sb_{0.12}$ versus temperature in magnetic fields of 0, 6, and 9 Tesla. The fact that the field is classically large in the temperature range of 5 – 150 K can be viewed by inspection of Figure 5. Since there is no change when increasing the field from 6 to 9 Tesla below 150 K, the high field limit has been reached and $\kappa_{carrier}$ has been completely suppressed. Above 150 K, there is a small change and the classical limit has not been achieved. As can be seen from Fig. 5, there is a large bipolar effect which is not eliminated by the magnetic field and results in the increase of the thermal conductivity. Therefore, extraction of Lorenz number using this method is only possible for temperatures below 150 K. Once the lattice and total thermal conductivities are measured, the electronic portion was calculated using Equation 1. Equation 2 can be rewritten as LT = $\kappa_{carrier}$ ρ where ρ is the zero field value for the electrical resistivity. Since in this case, the lattice portion is measured over a range of temperature, $\kappa_{carrier}$ ρ can be plotted versus temperature and the slope of the line will yield L for that temperature range. Figure 6 shows only the portion of the temperature range over which the plot is linear. At higher temperatures, above 150 K, the classical field approximation is no longer valid due to a drastic decrease in mobility as well as the onset of the bipolar contribution,[7] while at lower temperatures $\kappa_{lattice}$ dominates and therefore $\kappa_{total}$ is unaffected by magnetic field as can be seen in Figure 5. Fitting linearly as shown in Figure 6, gives the measured value for the Lorenz number to be $2.21\times10^{-8}$ $V^2K^{-2}$ in the temperature range 35-150 K; meaning L is constant over this range of temperature. Comparing with the values determined by Sharp et al.,[6] who obtained L = $2.31\times10^{-8}$ $V^2K^{-2}$ at 100 K, it can be seen that there is less than a 5% difference between the two measurements. When comparing values for the lattice portion of the thermal conductivity the measured value at 100 K yields 2.14 W/mK while the value determined in the literature is 2.19 W/mK. It should be noted that the grain sizes in both samples are of the same order of magnitude, with average grain sizes being roughly one and five microns for our sample and that of Sharp respectively.[6] Again, as in the MTR method, the measured values are not only reasonable, but within 5% of published values on the same material.

The same procedure was applied to all samples at 250 K. However the results were not as conclusive as those presented here. Different reasons for the ambiguity of the measurement include the onset of the bipolar effect in $Bi_{0.88}Sb_{0.12}$, as well as the mobility being drastically decreased at higher temperatures in all three samples, which has been seen previously.[1,5-8] This technique could prove useful in nanostructured materials where the lattice and electronic thermal conductivities become comparable.[1,5-9] Further investigation is required into higher temperature measurements as well as other types of materials[14,17] for which this technique can be useful; for example, magnetic fields up to 9 T had no effect on the thermal or electrical conductivity of nanostructured $FeSb_2$ and therefore the Lorenz number could not be determined experimentally.

## IV. CONCLUSION

Two methods for experimentally determining the Lorenz number are presented for nanopolycrystalline $Bi_{0.88}Sb_{0.12}$, $Cu_{0.01}Bi_2Te_{2.7}Se_{0.3}$, and $[Cu_{0.01}Bi_2Te_{2.7}Se_{0.3}]_{0.98}Ni_{0.02}$. Measured values of $Cu_{0.01}Bi_2Te_{2.7}Se_{0.3}$ and $[Cu_{0.01}Bi_2Te_{2.7}Se_{0.3}]_{0.98}Ni_{0.02}$ analyzed using equations 1-3 as well as equations 7-8 yield similar results and are close to calculated values using the single parabolic band model presented in equations 4-6. The measured values for $Bi_{0.88}Sb_{0.12}$ are the same as previously published results. Now that the two methods have been clearly demonstrated to work on these nanopolycrystalline alloys at a given temperature, it is possible to look at other materials as well as the temperature range for which this technique can be used. A systematic study can then be done of the temperature dependence of the Lorenz number for a given material, making it possible for more complex theoretical models to be verified within experimental error leading to more accurate determinations of the lattice portion of the thermal conductivity.

Acknowledgements: This work is supported by "Solid State Solar-Thermal Energy Conversion Center (S3TEC)", an Energy Frontier Research Center founded by the U.S. Department of Energy, Office of Science, Office of Basic Energy Science under award number DE-SC0001299/DE-FG02-09ER46577.

List of Figures and Captions

Figure 1: Thermal conductivity is plotted against electrical conductivity of $Cu_{0.01}Bi_2Te_{2.7}Se_{0.3}$ at 100 K with the magnetic field being varied from 0.8 T to 5 T. The slope of the linear fit provides the Lorenz number L = $2.16\times10^{-8}$ $V^2K^{-2}$ and the y-intercept gives $\kappa_{lattice}$ = 1.49 W/mK. The upper left inset plots the dependence of the total thermal conductivity on magnetic field. The lower right inset plots the dependence of electrical conductivity on magnetic field. Both the thermal and electrical conductivity varying with field can be fit using $\frac{aB^2}{1+cB^2}$ as shown in the insets.

Figure 2: Thermal conductivity is plotted against electrical conductivity of $[Cu_{0.01}Bi_2Te_{2.7}Se_{0.3}]_{0.98}Ni_{0.02}$ at 100 K with the magnetic field being varied from 0.8 T to 5 T. The slope of the linear fit provides the Lorenz number L = $2.33\times10^{-8}$ $V^2K^{-2}$ and the y-intercept gives $\kappa_{lattice}$ = 1.27 W/mK.

Figure 3: Electrical conductivity is plotted against magnetic field from 0.1-5 T and fit using equation 7. The electrical conductivity in zero field is used in order to determine the drift mobility, $\mu_d$ .

Figure 4: Thermal conductivity is plotted against magnetic field from 0.1-5 T and fit using equation 8 and $\mu_d$ from Figure 3. It is found that $\kappa_{lattice}$ = 1.29 W/mK.

Figure 5: Thermal conductivity is plotted against temperature at magnetic fields of 0, 6, and 9 T.

Figure 6: $\kappa_{carrier}$ ρ is plotted against temperature from 35-150 K. The black points represent the measured data while the red line is the linear fit. The slope of the linear fit provides the Lorenz number L = $2.21\times10^{-8}$ $V^2K^{-2}$ and the y-intercept gives $\kappa_{lattice}$ = 2.14 W/mK.

Figure 1

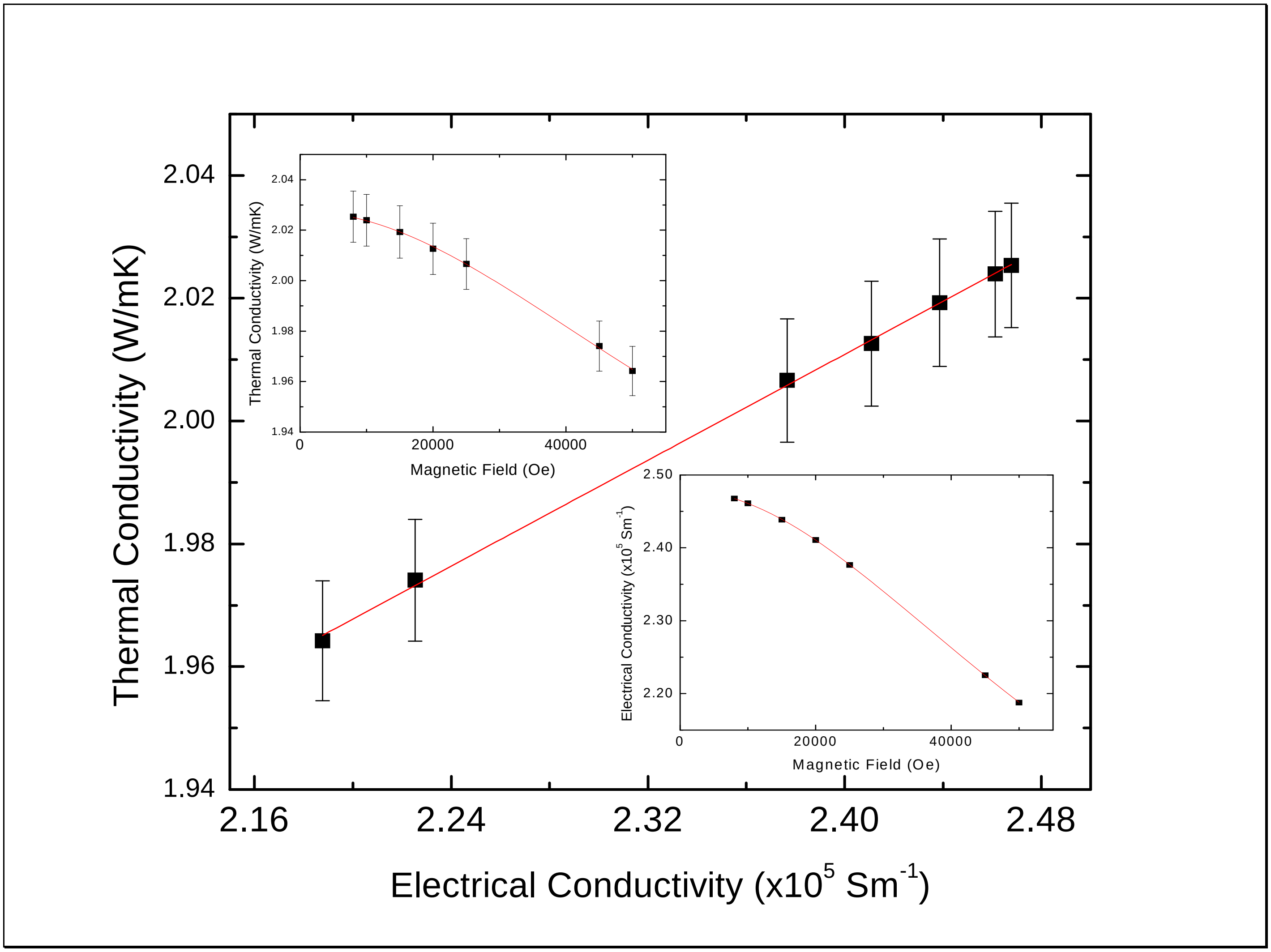

Figure 2

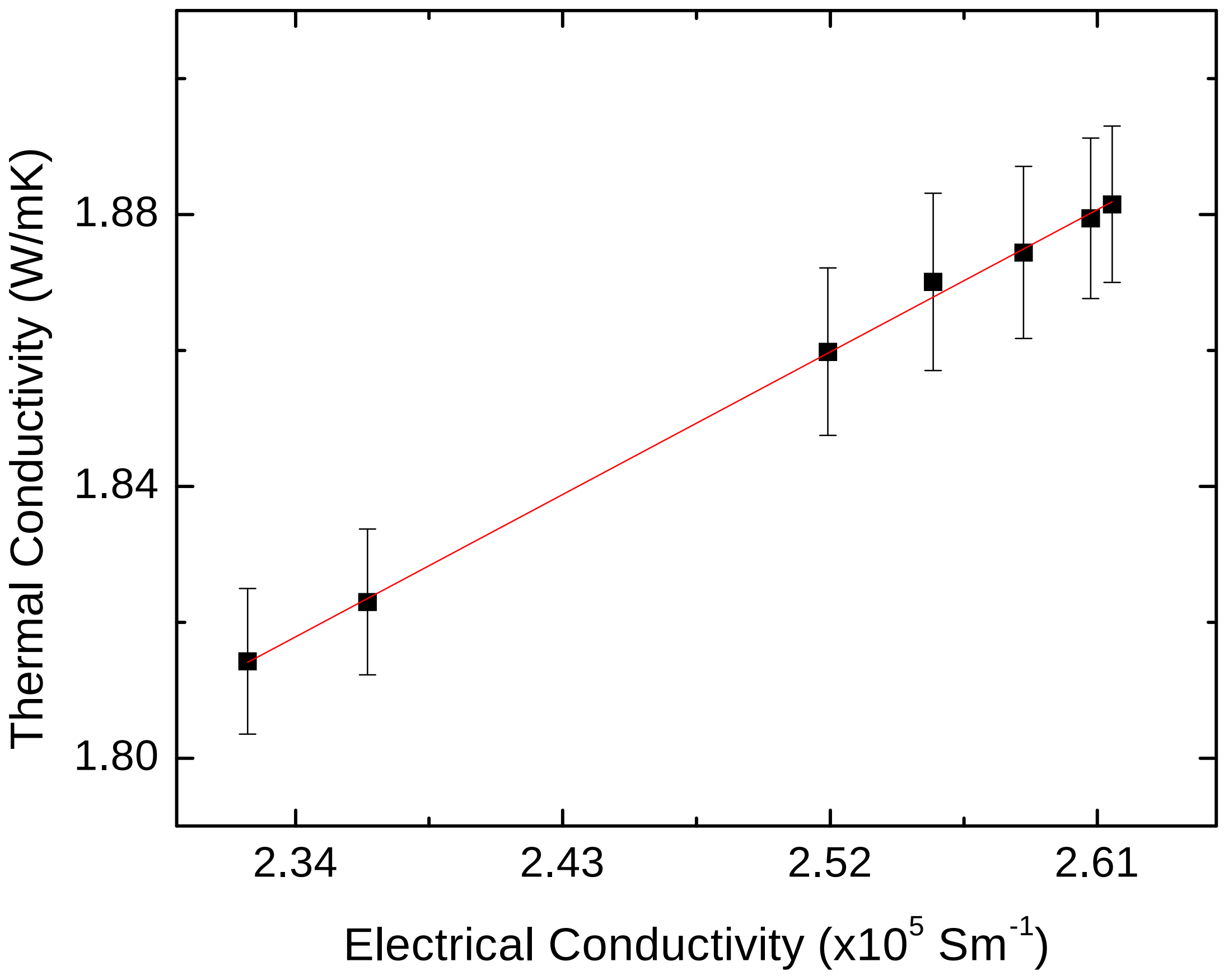

Figure 3

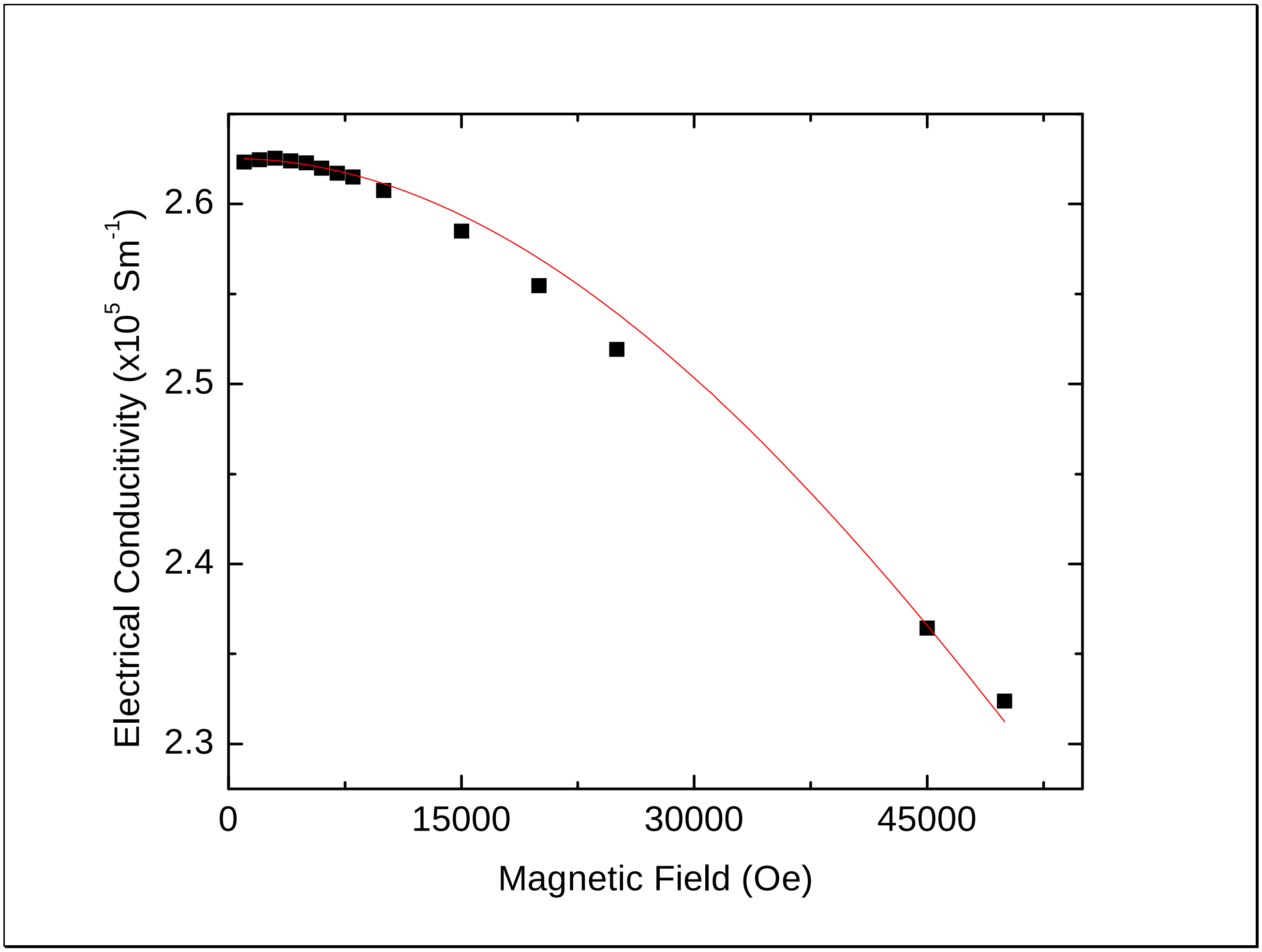

Figure 4

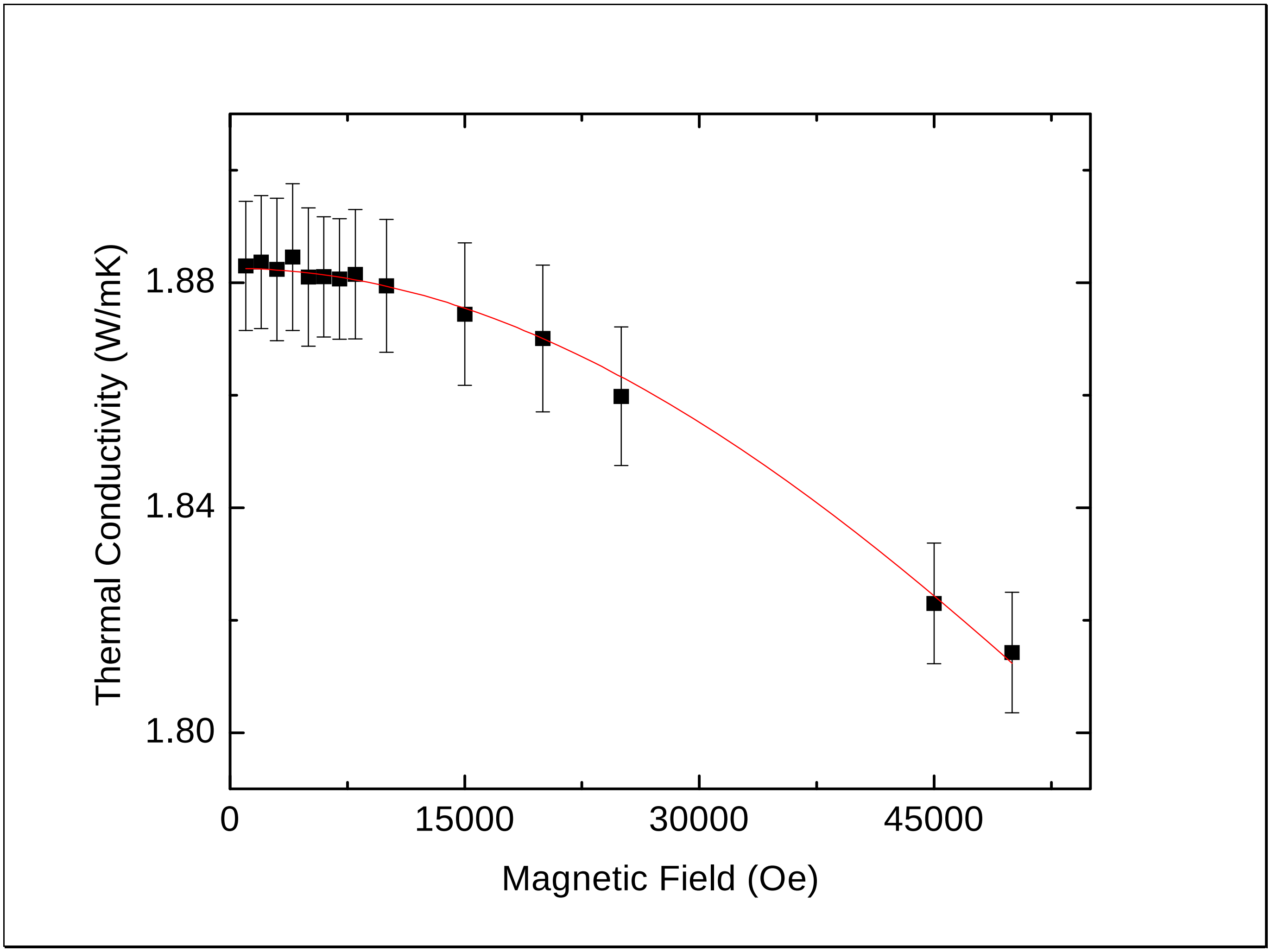

Figure 5

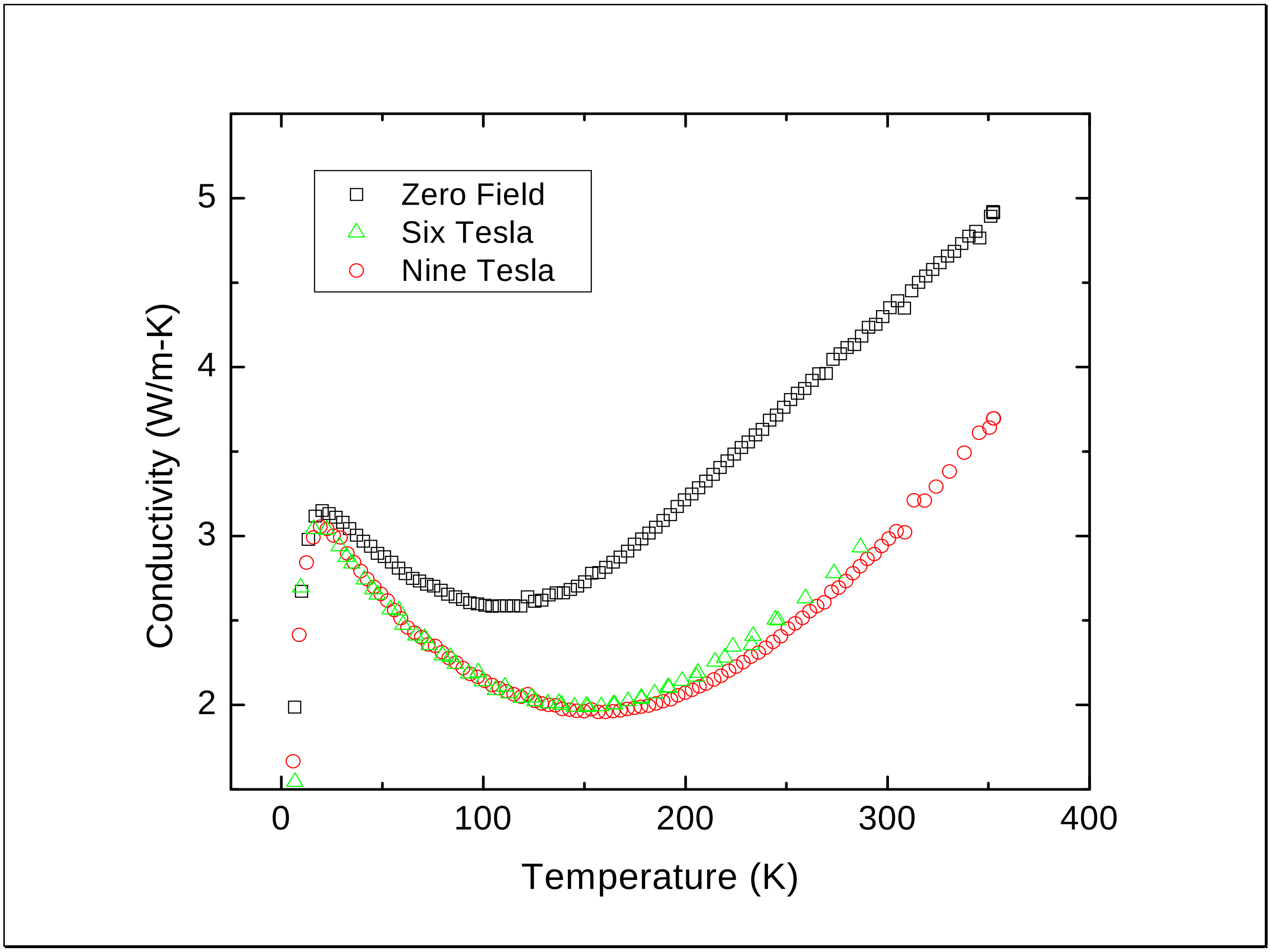

Figure 6

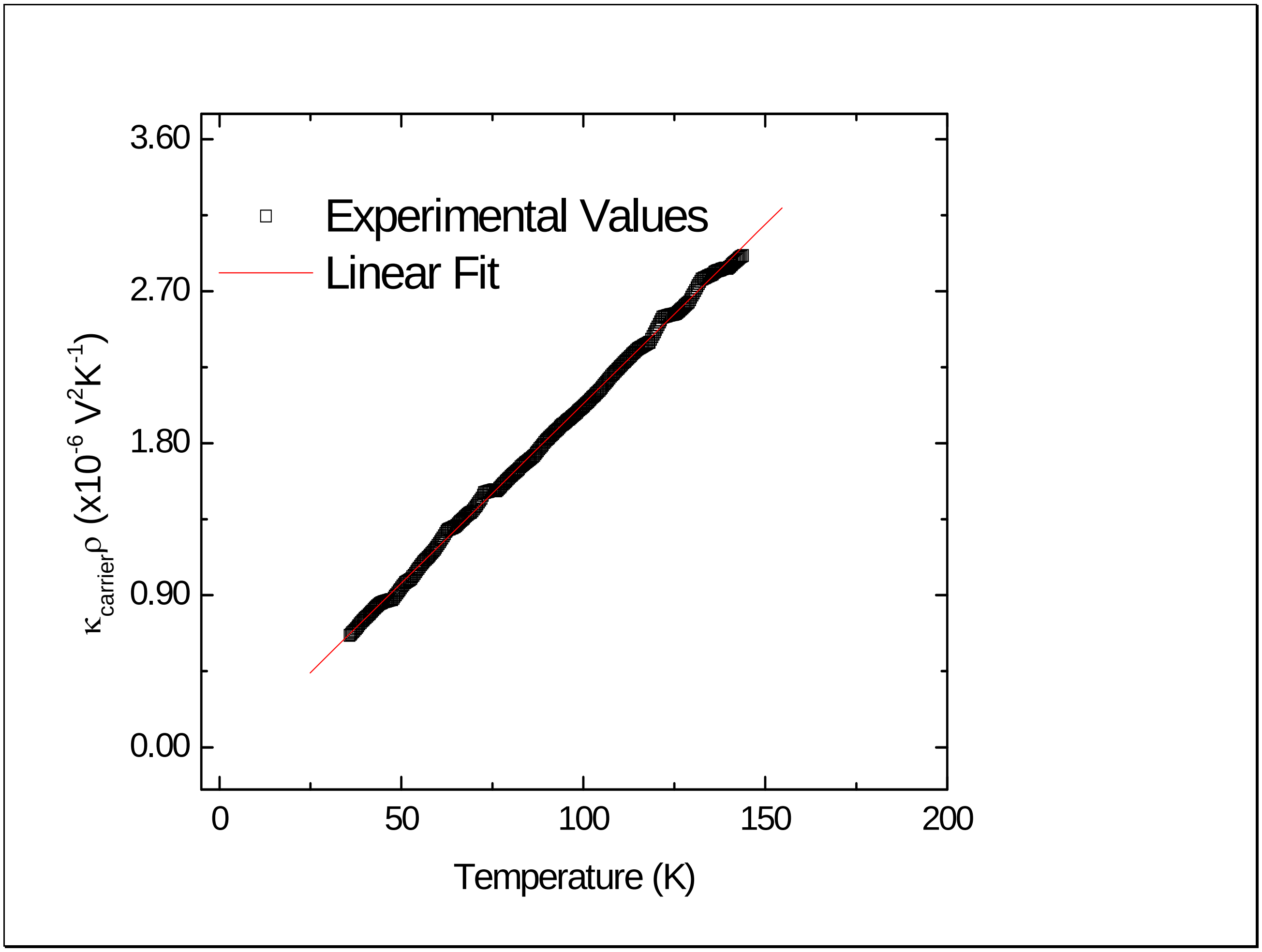